\documentclass[aps,prl,twocolumn,showpacs,preprintnumbers,amsmath,amssymb,10pt]{revtex4-1}

\usepackage{graphicx} 
\usepackage{pslatex} 
\usepackage{bm}       
\usepackage{textcomp} 
\usepackage{url}     

\newcommand{\vect}[1]{\boldsymbol{#1}}
\newcommand{\micron}{\hbox{\textmu}\text{m}} 

\begin{document}

\title{Determination of electric field, magnetic field, and electric current distributions of infrared optical antennas: A nano-optical vector network analyzer}

\author{Robert L. Olmon$^{1,2}$, Matthias Rang$^{2,4}$, Peter M. Krenz$^5$, Brian A. Lail$^6$, Laxmikant V. Saraf$^7$, Glenn D. Boreman$^5$, and Markus B. Raschke$^{2,3}$}%
\affiliation{%
$^1$Depts.~of Electrical Engineering, $^2$Chemistry, and $^3$Physics, University of Washington, Seattle, WA, 98195\\
$^4$Forschungsinstitut am Goetheanum, CH 4143 Dornach, Switzerland,\\
$^5$Center for Research and Education in Optics and Lasers (CREOL), University of Central Florida,~Orlando, FL 32816\\
$^6$Dept.~of Electrical and Computer Engineering, Florida Institute of Technology, Melbourne, FL 32901\\ 
$^7$Environmental Molecular Sciences Laboratory, Pacific Northwest National Laboratory, Richland, WA 99352
}%

\date{\today}


\begin{abstract}
In addition to the electric field $\boldsymbol{E}(\boldsymbol{r})$, the associated magnetic field $\boldsymbol{H}(\boldsymbol{r})$ and current density $\boldsymbol{J}(\boldsymbol{r})$ characterize any electromagnetic device, providing insight into antenna coupling and mutual impedance. 
We demonstrate the optical analogue of the radio frequency vector network analyzer implemented in interferometric homodyne scattering-type scanning near-field optical microscopy ({\it s}-SNOM) for obtaining $\boldsymbol{E}(\boldsymbol{r})$, $\boldsymbol{H}(\boldsymbol{r})$, and $\boldsymbol{J}(\boldsymbol{r})$.
The approach is generally applicable and demonstrated for the case of a linear coupled-dipole antenna in the mid-infrared. 
The determination of the underlying 3D vector electric near-field distribution $\boldsymbol{E}(\boldsymbol{r})$ with nanometer spatial resolution and full phase and amplitude information is enabled by the design of probe tips with selectivity with respect to $E_{\parallel}$ and $E_{\perp}$ fabricated by focused ion-beam milling and nano-CVD.
\end{abstract}

\pacs{PACS numbers: 78.67.-n, 68.37.Uv, 84.40.Ba, 73.20.Mf}


\maketitle

%
Optical antennas provide the ability to control and confine light on the nanoscale with applications including nanofocusing for field-enhanced spectroscopy and microscopy, coupling to surface plasmon polariton waveguides and optical nanocircuits, metamaterials, photodetectors, and thermal and molecular sensors \cite{schuck05, bharadwaj09, farahani05, zhang08, bozhevolnyi07, wenger08, alda05}. 
However, optical antenna design with desired functionality and matched impedance \cite{tang08, alu08, schnell09} has remained challenging compared to the radio frequency (RF) regime due to the lack of discrete circuit elements such as baluns and couplers. 
Instead, one relies on intrinsic optical transitions defined by free carrier, interband, intraband, and related polariton excitations with their geometric resonances giving rise to a high yet poorly understood sensitivity to these material properties and nanoscopic structural details.

As the primary source term  for the optical magnetic $\vect{H}(\vect{r})$ and electric $\vect{E}(\vect{r})$ vector fields of the antenna, the underlying current density distribution $\vect{J}(\vect{r})$ reflects the fundamental electrodynamic interaction and local coupling of antenna elements.
While $\vect{E}(\vect{r})$ and the weaker $\vect{H}(\vect{r})$ fields located outside the device structure only indirectly reveal the details of their microscopic origin via their spatial distribution, 
knowledge of the antenna current standing wave can provide direct and sensitive insight into impedance distribution, resonant frequency, or the coupling with neighboring antennas \cite{balanis, sundaramurthy05}.
Access to the current distribution is thus desired for optical antenna design and coupling to antenna loads, yet even with special nonlinear \cite{dadap99} and THz techniques \cite{seo07} direct current measurements with high spatial resolution have remained difficult experimentally. 

\begin{figure}
  \includegraphics[width=8.63cm]{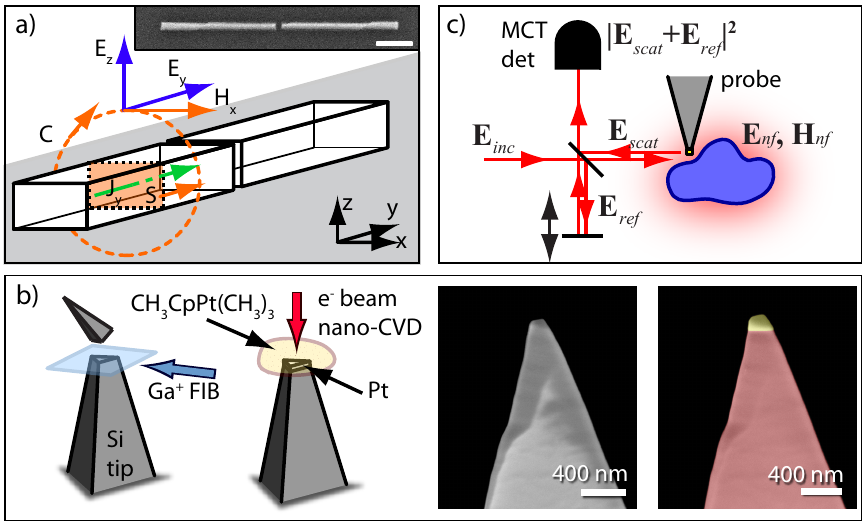} 
  \caption{Schematic of 
  antenna geometry (a) with current density $J(y)$ through cross sectional surface $\vect{S}$ and associated magnetic field $H_x(y,z)$ on contour $\vect{C}$ related to the measured curl of the electric field $\vect{E}(y,z)$ in the antenna $y$-$z$ mirror plane. Inset: SEM image of coupled antenna, scale bar 500 nm. 
(b) Pt point dipole probe antenna fabricated on a Si AFM tip by FIB and nano-CVD. 
SEM images of the tip before (left) and after (right) Pt deposition with Si (red) and Pt (yellow) regions highlighted.
(c) Schematic of amplitude-, phase-, and polarization-resolved interferometric homodyne 3D $\vect{E}$($\vect{r}$) vector near-field {\it s}-SNOM imaging.
}
\label{fig:setup}
\end{figure}

Here we demonstrate the determination of the conduction current density distribution $\vect{J}(\vect{r})$ and its associated magnetic vector field $\vect{H}(\vect{r})$ from measurement of the antenna 3D electric vector field $\vect{E}(\vect{r})$, taking advantage of the vector relationship in free space given by Faraday's Law $\vect{H}(\vect{r}) = \text{i}/(\omega\mu_0)\nabla\times\vect{E}(\vect{r})$ and Hall\'{e}n's integral equation relating in-plane $\vect{E}(\vect{r})$ to $\vect{J}(\vect{r})$~\cite{balanis}. 
This emphasizes the powerful implication that if $\vect{E}(\vect{r})$ is known with sufficient detail, then all electrodynamic parameters describing the optical response may be deduced.
We measure $\vect{E}(\vect{r})$ in the reactive near-field with high sensitivity and nanometer spatial resolution by a special implementation of scattering-type scanning near-field optical microscopy ({\it s}-SNOM) (see Fig.~\ref{fig:setup}a).
We derive $\vect{H}$($\vect{r}$) and $\vect{J}(\vect{r})$ for an infrared (IR) linear coupled dipole optical antenna resonant at 28.3~THz ($\lambda=10.6$~\micron), identifying the coupling between the antenna segments from details of the field and current distributions. 
With the ability to determine $\vect{E}(\vect{r})$, $\vect{H}(\vect{r})$, and $\vect{J}(\vect{r})$ with amplitude and phase with nanometer spatial resolution this work demonstrates the optical analogue of the RF vector network analyzer.

Although optical $\vect{H}$($\vect{r}$) can, in principle, be measured directly as shown recently using split-ring resonator probes \cite{burresi09}, 
the weaker Lorentz force associated with the magnetic light-matter interaction, compared to the Coulomb interaction, results in general in a higher detection sensitivity for $\vect{E}$($\vect{r}$), 
from which the associated $\vect{H}$($\vect{r}$) field can be derived in free space. 

A detailed understanding of IR optical antennas in terms of directional response, wavelength selectivity, and capture cross section is desirable for device applications such as high sensitivity thermal imaging, IR plasmonics, chemical sensing, or direct solar energy conversion.
With its mirror symmetry the coupled IR dipole antenna provides a well defined model system.
However, despite its simplicity, fundamental questions regarding antenna scaling behavior still remain open \cite{novotny07}, 
in particular in the mid-IR spectral range related to the transition between the low-energy Hagen-Rubens regime ($\omega\ll1/\tau_{Drude}$) characterized by high conductivity and low absorption, and the relaxation regime $1/\tau<\omega<\omega_{\text{pl}}$ with plasma frequency $\omega_{\text{pl}}$ \cite{dressel}. 
Yet with a Drude relaxation time $\tau_{\text{Drude}}\simeq30~$fs for Au corresponding to the optical cycle period of $\lambda \simeq 10~\micron$ radiation, devices operating in the mid-IR retain the ability to sustain surface plasmon polaritons, with enhanced propagation length \cite{jones09}.

{\it s}-SNOM was previously used to study 
only selected electric-field vector components of optical antennas \cite{olmon08, schnell09, yu07b}. 
Consideration of strong tip-sample coupling \cite{novotny08} or tip scattering anisotropy \cite{lee07, gersen07} 
using nanoparticle functionalized probe tips was found to be critical for measuring the full electric vector near-field distribution $\vect{E}$($\vect{r}$). 
Aperture probes have been used to measure the in-plane components of light propagating through photonic crystal waveguides~\cite{burresi09b}. 
For our generalized approach of vector resolved detection of $\vect{E}$($\vect{r})$ we engineer probe tips with defined polarization response and scattering sensitivity with regard to orthogonal 
$E_{\parallel}$ and $E_{\perp}$ 
field components. 
Following a theoretical design (see appendix), by combining focused ion beam (FIB) milling and electron beam assisted metalorganic chemical vapor deposition (nano-CVD) an off-resonant Pt point dipole probe is fabricated perpendicularly onto the apex of a commercial AFM Si cantilever tip as shown in Fig.~\ref{fig:setup}(b).
A low depolarization scattering, as critical for vector-resolved {\it s}-SNOM, of the resulting Pt-Si composite tip is verified experimentally with minimal depolarization of $\sim1~\%$.
Despite the 200~nm size triangular platelet Pt probe oriented under a slight tilt angle of a few degrees with respect to the sample plane, the {\it s}-SNOM spatial resolution is found to be dominated by the Pt prism corner dimensions 
of a few 10's of nm.

The resonant linear Au dimer antennas are fabricated on a Si substrate by electron beam lithography and lift-off. 
The nominal size of each antenna is $1.7~\micron\times110~\text{nm}$ with a height of 70 nm.
This length is chosen to correspond to the primary dipolar resonance with excitation at $\lambda=10.6~\micron$ considering the effective wavelength scaling as established previously \cite{olmon08,neubrech06,novotny07}.
They are arranged in collinear pairs to form coupled dimers with inter-antenna gaps of 80~nm \cite{olmon08}.

The general {\it s}-SNOM measurement scheme is shown in Fig.~\ref{fig:setup}(c).
With the object under far-field excitation of $\vect{E}_{{\text{inc}}}$ of defined polarization, the tip-scattered near-field signal
$\vect{E}_{{\text{scat}}}\propto\vect{E}_{{\text{nf}}}$ is collected in a collinear backscattering geometry,  with interferometric homodyne amplification for amplitude, phase, and polarization resolved detection \cite{keilmann04, jones09, bek06} with 
$S_{{\text{det}}}=|\vect{E}_{{\text{scat}}}+\vect{E}_{{\text{ref}}}|^2$.
Specifically for this work the IR dipole antennas are excited using a CO$_2$ laser 
polarized parallel with respect to the antenna axis (s-pol), incident at an angle of 60$^\circ$ with respect to the surface normal, using a Cassegrain objective ($\text{NA}=0.5$) \cite{jones09,olmon08}.
With in-plane excitation, tip-sample coupling is minimal as can be seen from image dipole theory. 

The spatial near-field distributions for s- and p-polarized detection
are measured in lift mode via a series of line scans.
With the lock-in amplifier referenced to the second-harmonic of the tip dither frequency for far-field background suppression \cite{keilmann04}, the signal is related to the intensity gradient of the optical near-field in the $z$-direction \cite{jones09}.
The field amplitude for each field component is then obtained numerically from the integral of the detected field gradient with integration constant combined into a background term.

%
\begin{figure*}
  \includegraphics[width=17.856cm]{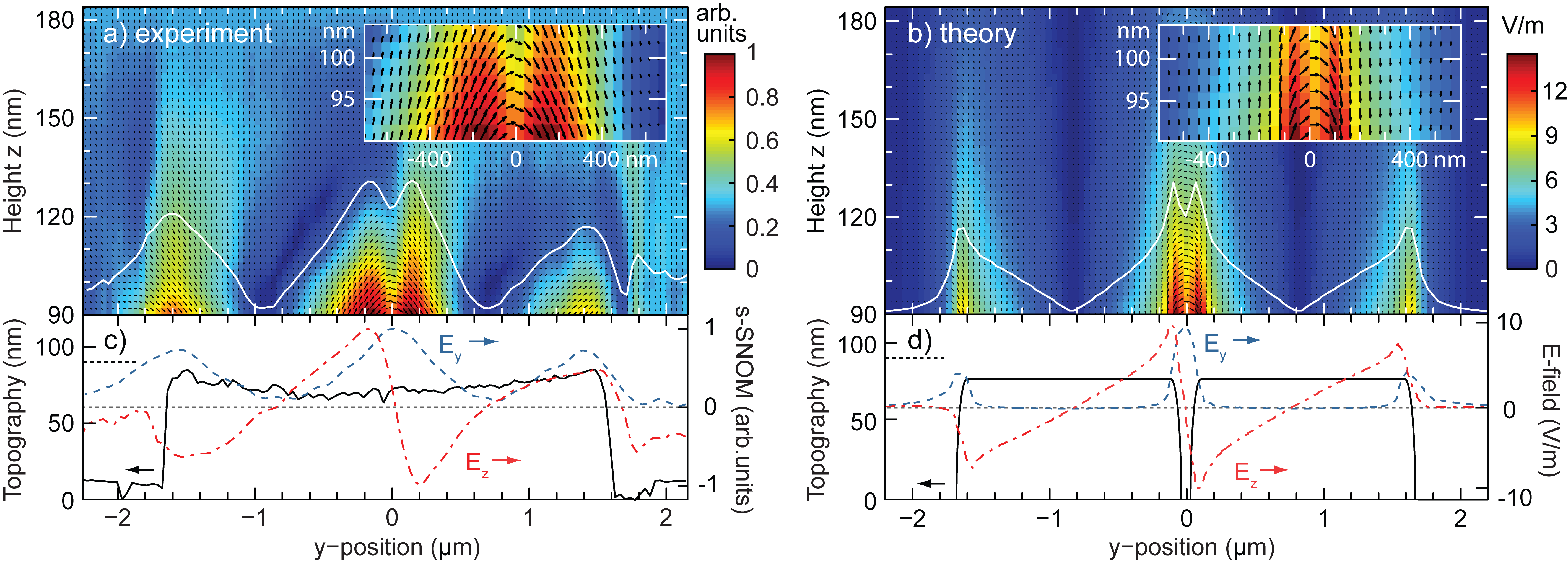}
  \caption{Measured (a,c) and theoretical (b,d) 
   near-field vector distribution of linear Au IR optical dimer antenna: (a) $\vect{E}(y,z)$ in the $y$-$z$ plane and (b) corresponding theory. Field magnitude is given by vector length and color map. A line scan (white line) for $|\vect{E}(y,z)|$ at $z=90$~nm is shown. Inset: Close-up views of the near-field vector distribution near $y=0$.
  Line scans (c,d) show topography and measured E-field components E$_y$ and E$_z$ at $z=90$ nm above the antenna surface. The 
  80 nm gap is not fully resolved in topography due to the probe size.}
    \label{fig:nf}
\end{figure*} 
Figure \ref{fig:nf} shows the resulting 2D $\vect{E}$($y,z$) near-field map in the $x=0$ plane above the dimer antenna (a) (inset, close-up view). 
The data are acquired with sampling steps of $42~\text{nm}\times2~\text{nm}$ (corresponding to $128\times48$~pixels) in the $y$- and $z$-directions, respectively.
The field vectors at each point indicate the direction and relative amplitude (also given by color code) of $|E_{{\text{nf}}}|=({E_{\text y}^2+E_{\text z}^2})^{1/2}$.
For comparison, a simulation with the antennas modeled as Au half cylinders terminated by quarter spheres on a Si substrate is shown (b).
The somewhat wider lateral extent of the experimental field distribution is likely due to signal convolution with the finite tip apex, imperfections in the antenna geometry, and surface roughness locally enhancing the fields near the Au surface. 

Line scans of topography and the two individual field components $E_{\text{y}}$ and $E_{\text{z}}$ at $z=90~\text{nm}$ are shown for experiment (c) and theory (d).
Three distinct regions of high field magnitude are seen in $E_{\text{y}}$, corresponding to
the locations of high charge concentration associated with the gap region and outer antenna terminals.
These three regions oscillate in phase, in contrast to the alternating phase changes of $\pi$ of the associated $E_{\text{z}}$ field \cite{olmon08,jones09}.
The $E_{\text{y}}$-dominated gap field is characterized by a large homogeneous field enhancement due to the antenna coupling, with small or zero $E_{\text{z}}$ component due to phase reversal 
in the gap. 
The total field $|E_{{\text{nf}}}|$ at $z=90$~nm is shown overlaid in Figs.~\ref{fig:nf}(a) and (b) (white line).
In contrast to earlier studies with unspecified polarization detection \cite{yu07b}, it is evident that 
the global field maximum occurs just outside the gap 
at the metal edge
where the field lines converge associated with the sharp curvature. 
Field enhancement, estimated from comparison with off-resonant signal levels 
is $15\pm5$ at the gap and $11\pm3$ at the outer terminals for $E_{\text{y}}$, 
and $32\pm18$ near the gap and $17\pm10$ at the outer terminals for $E_{\text{z}}$ 
with spatial $1/\text{e}$ decay length of 38~nm and 33~nm, respectively, 
in good agreement with theory and related experiments \cite{neubrech06, jones09}.

{\it Recovery of H(r) and J(r)}. 
With the plane wave excitation with wave vector approximately perpendicular to the $y$-$z$ mirror plane of the coupled dipole antenna 
where $E_{\text{x}}=0$\,V/m, 
the measured 2D distribution of $E_{\text{y}}$ and $E_{\text{z}}$ is sufficient to fully characterize the antenna response. 
Applying Faraday's Law the corresponding magnetic field is obtained from the curl of the electric field as $H_{\text{x}}(y,z)=\text{i}(\partial E_{\text{z}}/\partial y - \partial E_{\text{y}}/\partial z)/(\omega \mu_0)$ (for procedural details see appendix). 
\begin{figure}
  \includegraphics[width=8.63 cm]{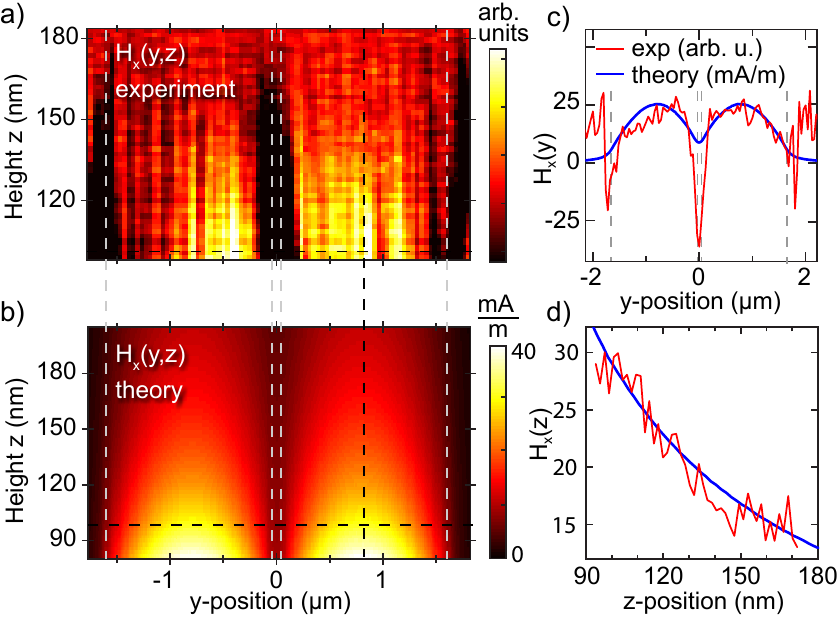}
  \caption{Magnetic field $H_{\text{x}}$($y$,$z$) above coupled dipole antennas (position indicated by gray dashed lines) derived from the the electric near-field distribution $\vect{E}$($\vect{r}$), experiment (a) and theory (b). 
Horizontal variation of $H_{\text{x}}$($y$) (c) and vertical distance dependence of $H_{\text{x}}$($z$) (d) (black dashed lines in a and b).
Due to the antenna symmetry, $H_{\text{x}}$($y$,$z$) in the $y$-$z$ mirror plane represents the full magnetic antenna response.}
\label{fig:Hx}
\end{figure}
Figure \ref{fig:Hx}(a) shows the resulting $H_{\text{x}}(y,z)$ distribution obtained from the measured $\vect{E}$($y,z$) field of Fig.~\ref{fig:nf}(a). 
Comparison with theory (b) shows good agreement with the characteristic central $H_{\text{x}}(y,z)$ maximum for each dipole, the corresponding minima at the extremities, and the decay length, as seen in the respective line averages of the magnetic field as a function of lateral position $y$ along the rod axis at height $z=100~\text{nm}$ (c), and along $z$ over the center of a single rod (d). 
A magnetic field enhancement $|H_{nf}|/|H_{inc}|$ of $\sim16$ can be estimated with the incident field approximated by $|H_x(z=180~\text{nm})|$ normalized to 1. This agrees with the corresponding value of $\sim15$ from theory considering $|H_{inc}|=2.7$~mA/m ($\equiv |E_{inc}| = 1$V/m).
The results emphasize the requirement for accurate and high resolution near-field $\vect{E}$($\vect{r}$) data, since the magnetic field determination relies on the difference between orthogonal field gradients, making it very sensitive to
noise, systematic errors in the detected signal, and depolarized scattering especially at the center of each  antenna segment where $\partial $E$_{\text{y}}/\partial z$ is small. 

\begin{figure}[tb]
  \includegraphics[width=8.63 cm]{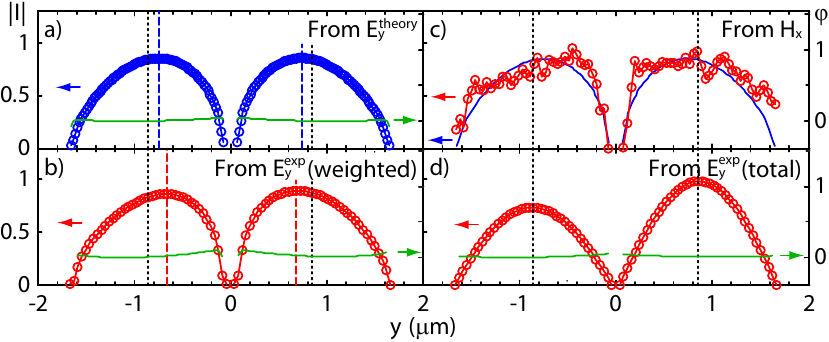}
  \caption{Conduction current distribution $I(y)$ along antenna dimer 
  reconstructed from the in-plane electric vector-field $E_y$, from theory (a) and experiment (b, weighted and d, total) with amplitude $|I|$ and phase $\phi$~[radians].
  A shift in current maximum (dashed line in a and b) of 100-200 nm with respect to geometric center (dotted lines) results from capacitive gap coupling.
  $I(y)$ calculated from $H_x$ (c) is shown versus the result from $E_y^{theory}$ for comparison (solid line).
  }
\label{fig:current}
\end{figure}

The antenna conduction current $I(y)$ shown in Fig.~\ref{fig:current} is retrieved in magnitude and phase $\phi$ from both the theoretical (a) and experimental (b and d) $E_y(z=90{\text{ nm}})$ (see Fig.~\ref{fig:nf}) using the method of moments to solve Hall\'en's integral equation with a pulse basis and point-matching \cite{orfanidis}. 
With the antenna radius $a\ll\lambda$, $\vect{J}$($\vect{r}$) is approximated as a linear current distribution with $I=0$ at the antenna terminals.
To compensate for the imperfections in the measurement due to surface roughness and related local-field enhancement at the Au surface with the high sensitivity of the moment method to field localization, the in-plane field is deemphasized along the antenna surface while the magnitude at the poles is maintained (see appendix for details). 
An alternative method of current retrieval using the relationship $\vect{J}(\vect{r})=\vect{\nabla}\times\vect{H}(\vect{r})$ at the Au/air interface (c) also shows good qualitative agreement with theory, yet relying on multiple derivatives of $\vect{E}$($\vect{r}$) it is very sensitive to noise as discussed above.

The inter-dimer antenna coupling via Coulomb interaction across the gap, with its greater electric field due to the high polarization gradient, increases the instantaneous current with maxima shifted towards the gap (Figs.~4a and b). 
This coupling is a manifestation of the mutual impedance between the interacting dipoles, which gives rise to a  
red  shift of the resonant frequency with increasing coupling strength \cite{olmon08, schnell09}.
The coupling leads to the higher $|\vect{E}_{nf}|$ near the gap region (Fig.~\ref{fig:nf}), and the off-center shift of the maximum of $H_{\text{x}}$ towards the gap (Figs.~\ref{fig:Hx}a and c), 
but it is most evident in the current distribution, with its distinct shift of $\sim100-200$ nm, expected to increase further for smaller gap spacing.

Current density distribution measurements in general can provide insight into the degree of antenna resonant mismatch.
Sustained current oscillations with well-defined maxima and minima indicate that the structure is resonantly matched to the effective wavelength, dependent on material and aspect ratio \cite{bharadwaj09}, as is the case for the antenna shown in Fig. 4.
In contrast, current reflections at poorly matched junctions to, e.g. sensors or waveguides, or resonance shifts due to strong mutual impedance between radiatively coupled antennas can result in drastically altered current distributions requiring adjustments in the antenna design. 
It can thus serve as a sensitive probe 
for optimization of optical antennas with impedance matched sensors or controlled field enhancement.

In conventional probing of RF devices, the magnitude and phase of the potential at the input of a scanning probe antenna (typically a small dipole or loop) is measured using a vector network analyzer or a spectrum analyzer with a known reference frequency with the probe acting as the receiver in transmission ($S_{21}$) mode~\cite{baudry07}.
As demonstrated here, the {\it s}-SNOM-based implementation with specifically engineered probe tips with respect to their polarization selective scattering together with the interferometric homodyne signal detection taking the place of the RF receiving and antenna reference field, represents the optical analogue of the RF vector network analyzer.

This approach has allowed us to address open questions regarding the details of confinement and spatial distribution of the vector electric field associated with the gap 
of coupled IR antennas~\cite{yu07b} as well as to identify the antenna source current distribution at infrared frequencies
~\cite{sundaramurthy05}.
For the case of linear dipole antennas, a single excitation/detection pathway probing of the two near-field components perpendicular to the incident $\vect{k}$-vector has been sufficient for the complete antenna characterization.
Generalization of the approach using two orthogonal optical pathways with collinear excitation and detection in each affords vector-resolved detection in full 3D, enabling $\vect{E}$, $\vect{H}$, and $\vect{J}$ measurement for arbitrary optical antenna geometries, thus providing an ideal tool for optical antenna characterization, analysis, and design.

Funding from the National Science Foundation (NSF CAREER Grant CHE 0748226 and IGERT) is greatfully acknowledged as is support from the Environmental Molecular Sciences Laboratory at Pacific Northwest National Laboratory.

\section{Appendix}
The performance of the vector near-field imaging method presented above is sensitive to the properties of the probe tip.
Here we describe further details of the fabrication process and results of simulations regarding tip scattering with and without the Pt probe antenna.
We also provide further procedural details regarding the vector field measurement with respect to individual field components and spatial resolution.
Last, discussion of the method for magnetic field and current density determination is provided.

\subsection{Probe Fabrication}
The very tip apex of a commercial atomic force microscope (AFM) Si probe tip (AdvancedTEC NC, $\rho=0.01\sim0.025\,\Omega\cdot$cm) 
 is cut off using a Ga$^+$ focused ion beam (FIB) (FEI Helios 600, Ga$^+$ current 90 pA, beam energy 30 keV) to create a well defined $\sim$200 nm wide plateau \cite{giannuzzi05}. 
The plateau is oriented at an angle of $60^{\circ}$ with respect to the tip axis as required by the {\it s}-SNOM configuration to ensure that its orientation is parallel with respect to the scanning plane.
A $\sim50$~nm thick platelet of Pt is subsequently deposited onto the plateau using {\it in situ} electron beam assisted nanoscale chemical vapor deposition (nano-CVD) with a CH$_3$CpPt(CH$_3$)$_3$ (trimethyl-platinum-methylcyclopentadienyl) precursor as the platinum source (electron beam  current $\sim$100 pA) \cite{vandorp08}.
This platelet probe antenna, with dimensions of $\sim\lambda/50$ is sub-resonant compared to the excitation wavelength of $\lambda = 10.6~\micron$ used in the experiment.
It can be considered a low-power Rayleigh point dipole emitter, thus minimally perturbing the antenna near-field to be investigated, similar to the use of a bare Si tip \cite{schnell10}.
Moreover, with antenna illumination perpendicular to the tip axis, thus resonantly exciting the sample but not the probe tip, undesired coupling effects between the tip and the sample are expected to be minimal for both detection polarization configurations.
This is in contrast to a tip-parallel illumination and detection configuration where strong tip-sample dipole-dipole coupling can distort the intrinsic antenna field thus making the interpretation of the recovered signal difficult \cite{esteban08}.

\subsection{Tip Simulations}
To aid in the probe design, the optical response characteristics of the original Si bulk probe and the modified nanocomposite Pt-Si probe are simulated based on the finite element method (HFSS, Ansoft LLC).
The bulk tip is modeled as a Si-cone oriented in the $z$-direction with a full taper angle of $20^{\circ}$, terminated by a hemispherical apex with 10 nm radius.
For the modified tip, the cone is truncated and capped with a cylindrical Pt layer in the $x$-$y$ plane, 25 nm thick and 100 nm in diameter.
A plane wave is incident propagating in the $y$-direction with a field strength of 1 V/m, polarized in either the $x$- or the $z$-direction.
It should be mentioned that the tip used in the experiment is covered by a native SiO$_{\text {2}}$ surface layer \cite{haefliger04}.
That thin layer, neglected in the simulation, is not expected to alter the general conclusions drawn from the simulation.

\begin{figure}
  \includegraphics[width=8.67cm]{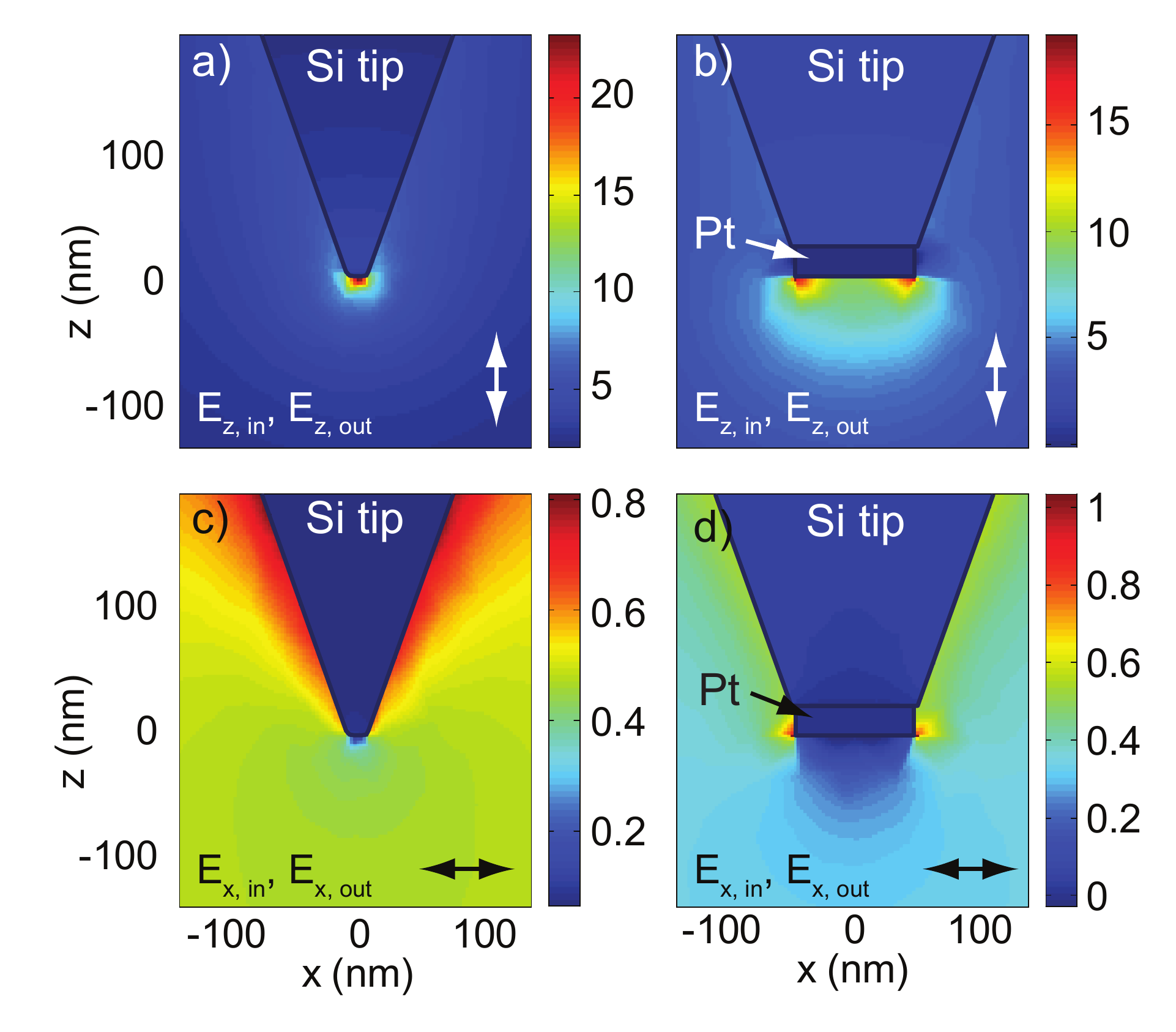}
  \caption{The simulated optical electric field response $|\vect{E}_{\text {i}}|$ (${\text {i}}=x,z$) for parallel (a and b) and perpendicular (c and d) illumination ($\lambda=10.6~\micron$) with respect to the tip axis for a Si bulk tip (a and c) and the modified probe tip (b and d) given in V/m. The image is shown at the time in the optical cycle of highest field amplitude. A dipolar response is clearly seen in the modified tip.}
  \label{fig:theory}
\end{figure}

Figure \ref{fig:theory} shows the expected optical response for parallel (a and b) and perpendicular (c and d) illumination with respect to the tip axis.
While the unmodified bulk tip shows a strong non-plasmonic antenna-type response under tip-parallel excitation with an enhancement factor of 24 at the apex (a), its response under tip-perpendicular illumination is negligible (c).
The bulk tip is thus largely insensitive with respect to the perpendicular vector component $E_{\perp}$. 
On the other hand, the Pt-terminated tip produces an enhanced field in response to both perpendicular (d) and parallel polarization (b) with the latter attenuated by only 17\% compared to the bulk tip thus giving rise to increased scattering via the induced dipole at the apex subject to the $E_{\perp}$ near-field component of the antenna being probed.

\subsection{Field Measurement}
To determine the vector orientation, $E_{\text {y}}$ and $E_{\text {z}}$ are measured independently with their relative scaling calibrated against theory. 
A carefully balanced weak homodyne reference field enables the assignment of the near-field phase at each sampling point. 

The relative temporal phase of the detected near-field can be adjusted from 0 to $2\pi$ by moving the reference mirror.
Spatial variation in phase can be understood as the direction of the measured near-field vector at each point with respect to the direction of the reference field vector at a specific time.
For the measurements shown, the relative phase was adjusted for maximum enhancement.
The spatial phase variation is observed separately for the two measured near-field components.
While for the $E_{\text {z}}$ component, a $\pi$-phase change is observed between the endpoints of each antenna, 
the $E_{\text {y}}$ field enhancements are all positive with respect to the near-field on the surrounding Si, indicating that they are in phase with each other, as expected for coupled dipoles.
The combination of the two field components yields the complete vector field with the relative amplitude between $E_{\text {y}}$ and $E_{\text {z}}$ established from theory, as its determination is a difficult task in general, requiring an absolute calibration of the experimental setup and tip response with respect to scattering of the confined near-field excitation.

Though the modified probe tip apex has a diameter of $\sim200$~nm, a spatial resolution of up to $\sim40$~nm as been obtained in the near-field images. 
With the probe tip under a slight tilt, the near-sample edge of the triangular Pt platelet ultimately defines the resolution and sensitivity.
With a tilt angle of just a few degrees, depolarization of the scattered signal remains negligible.

\subsection{Calculation of current from electric field data}

In general, the electric vector field $\vect{E}$ and magnetic field $\vect{H}$ of the time-varying optical electromagnetic wave are related by Faraday's Law.
\begin{equation}
\label{eqn:faraday}
\vect{H}= 
-\frac{\text{i}}{\omega\mu_0}\frac{\partial \vect{B}}{\partial t}=
\frac{\text{i}}{\omega\mu_0}\vect{\nabla}\times\vect{E}.
\end{equation}
Here we use the constitutive relation $\vect{B}=\mu_0\vect{H}$ and the fact that the time derivative of a time-harmonic wave can be represented instead as a multiplication by $\text{i}\omega$ where $\omega$ is the frequency of the wave.

Using the geometry from Fig.~1, the antenna near-field is measured in the mirror plane defined by $x=0$. 
This simplifies the numerical curl operation.
In this geometry, the magnetic field consists of only a $x$-component produced from the spatial derivatives of the electric field in the $y$- and $z$-directions, and Eq.~\ref{eqn:faraday} simplifies to
\begin{equation}
\label{eqn:magfield}
H_x=
\frac{\text{i}}{\omega\mu_0}\left(\frac{\partial E_{\text {z}}}{\partial y}-\frac{\partial E_{\text {y}}}{\partial z}\right).
\end{equation}

The electric field in space $\vect{E}(\vect{r})$ and the associated current density $\vect{J}(\vect{r'})$ of a conductor are related through the magnetic vector potential $\vect{A}(\vect{r})$ with $\vect{B}(\vect{r})=\vect{\nabla}\times\vect{A}(\vect{r})$.
Approximating $\vect{J}$ as a line current, $I(y')$ is determined directly by the measured electric near-field antenna-parallel component $E_y$ and can be obtained as the solution to Hall\'en's integral equation
\begin{equation}
\frac{\mu_0}{4\pi}\int^{l/2}_{-l/2}{I(y')G(y-y')dy'=\text{i}\omega\mu_0\epsilon_0(\partial_y^2+k^2)^{-1}E_y(y)}
\end{equation}
with antenna length $l$ in wavelengths, $k^2=\omega^2/\mu_0\epsilon_0$, and kernel $G(y-y')$ related to the geometry with $R=\sqrt{(y-y')^2 + z^2}$ the distance from each point of $I(y')$ to each point of $E_y(y,z)$~\cite{balanis, orfanidis}.
This equation is solved numerically by the method of moments with a pulse basis and point matching using a MATLAB code following Ref.~\cite{orfanidis}.

\begin{figure}
  \includegraphics[width=8.67cm]{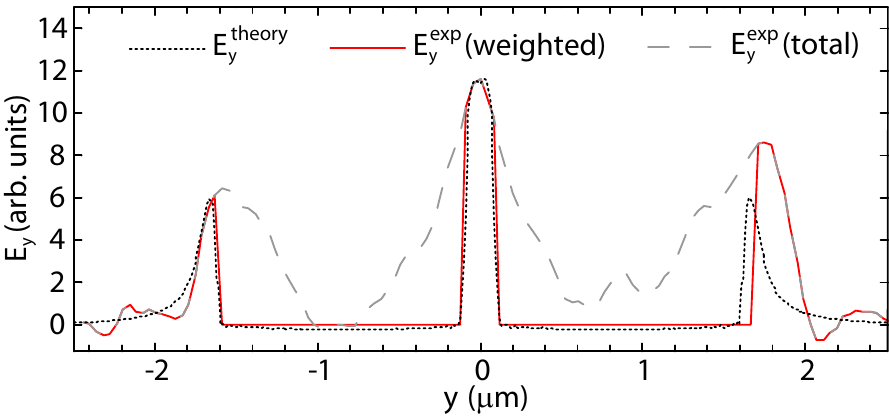}
  \caption{The total $E_y$ is weighted to preserve the field amplitudes at the poles but to deemphasize the surface field as necessitated by field broadening attributed to interactions between the probe tip and surface roughness on the antenna.}
  \label{fig:weighting}
\end{figure}

Application and comparison of the integral equation method using the theoretical $E_y$ field indicates a numerical challenge associated with the surface roughness on the antenna and its associated extraneous field-enhancement that gives rise to the broadening of the measured $E_y$ field distribution observed.
The resulting $I(y)$ (Fig. 4~(d)), while exhibiting the sinusoidal distribution as expected underepresents the antenna coupling showing indiscernible current maxima shifts due to the sensitivity of the method of moments to the roughness induced broadened $E_y$ distribution.
This necessitates the use of a weighting function to deemphasize the in-plane field along the Au surface while maintaining the strong driving fields at the poles.
Figure \ref{fig:weighting} shows $E_y^{theory}$, $E_y^{exp}$(total), and $E_y^{exp}$(weighted) from theory and experiment respectively, which were used to produce the currents shown in Fig. 4.
Deemphasis of the field on the surface retains the main features at the antenna terminals that agree well with theory in terms of amplitude, phase, and spatial distribution.

Alternatively, the conduction current can be approximated from the magnetic field data.
Near the conductor surface, in the absence of displacement current, the underlying source current density is $\vect{J}=\vect{\nabla}\times\vect{H}$.
With $H_x$ given above, considering that the current is largely one-dimensional and dominated by the $y$-component for the linear antenna geometry studied, the curl equation can be simplified to 
\begin{equation}
J(y) = -\frac{\partial}{\partial z}H_x(y,z). 
\end{equation} 
This method, however, is not preferred over the integral equation method, since it involves several derivatives from the original $\vect{E}$ data, resulting in amplification of noise, and it neglects contributions from any displacement current.
Nevertheless, it does provide an additional check on the expected $I(y)$ distribution as shown in Fig. 4(c).

Thus if $\vect{E}$ is known in sufficient detail, one may calculate the associated $\vect{H}$ and $\vect{J}$.
Though simplified here for the case of a linear antenna geometry, these operations are general and can readily be extended for the determination of magnetic field and current from 3D near-field data for arbitrary antenna geometries.


\clearpage

\clearpage

\end{document}